\documentclass{article}

\usepackage{graphicx}
\usepackage{dcolumn}
\usepackage{bm}
\usepackage{amsmath}
\usepackage{amssymb}

\newcommand{\cact}[1]{[{\rm C}_{#1}]}
\newcommand{\catot}[1]{[{\rm C}_{#1}]_{\rm T} }
\newcommand{\cinact}[1]{[\widetilde{{\rm C}}_{#1}]}
\newcommand{\ac}[1]{[{\rm AC}_{#1}]}
\newcommand{\acon}{[{\rm A}]}
\newcommand{\at}{\acon_{\rm T}}
\newcommand{\plc}[1]{{\rm C}_{#1}}

\begin{document}

\begin{center}
{\bf \large An allosteric model of circadian KaiC phosphorylation}\\
{\bf Supporting Information}\\ Jeroen S. van Zon, David K. Lubensky,
Pim R. H. Altena,\\ and Pieter Rein ten Wolde
\end{center}
\vspace*{0.5cm}

\tableofcontents

\vspace*{1cm}
\noindent
In this {\em Supporting Information}, we provide background
information on our model of the {\em in vitro} Kai system and the
calculations that we have performed. We will closely follow the
outline of the main text.

\section{Allosteric Model}
In this section, we discuss in more detail the allosteric model of
section I of the main text. We first present a statistical-mechanical
description of the allosteric model. This allows us to describe the
thermodynamics of the phosphorylation cycle. We then present a model
based on concepts of transition state theory that allows us to
describe the {\em dynamics} of the phosphorylation cycle, in particular the
dynamics of the conformational transtions. Lastly, we briefly discuss
how we have performed the simulations on the model in Fig.1B of the
main text, the results of which are shown in Fig.1C of the main text.

\subsection{Thermodynamics: A statistical-mechanical model}
 The allosteric model relies on the following key assumptions:
\begin{enumerate}
\item Each of the $N = 6$ monomers of a hexamer is either in an
active or in an inactive conformational state.
\item The conformations of the monomers are strongly coupled, such  that all
monomers of a hexamer are in the same conformational state at all
times.
\item Both in the active and inactive state, monomers can be
  (de)phosphorylated and (un)bind nucleotides. We assume that
  nucleotides have a higher affinity for monomers in the active state
  than for those in the inactive state. Consequently, nucleotide
  binding enhances the stability of the active state with respect to
  the inactive one. In contrast, we assume that phosphorylation favors
  the inactive state.
\item Nucleotide exchange is faster than phosphorylation and therefore
  in thermodynamic equilibrium on the time scale of phosphorylation.
\end{enumerate}
The model makes the following further assumptions that are of secondary importance:
\begin{enumerate}
\item Each monomer has two phosphorylation states, phosphorylated and
unphosphorylated. Each hexamer thus has $N=6$ phosphorylation sites.
\item
Phosphorylation of the different monomers of a hexamer occurs
sequentially around the hexamer.
\item
The unphosphorylated monomers can bind ATP, while the phosphorylated
monomers can bind ADP.
\end{enumerate}

This leads to the following partition function for a hexamer that it
is in a conformational state $\alpha$, at phosphorylation level $p$,
with $q$ ATP  and $r$ ADP molecules bound:
\begin{eqnarray}
Z^{\alpha} (p,q,r) &=& e^{-\beta N E^{\alpha}_{\rm m}} \binom{N-p}{q}\left({\rm [ATP]}/K_{\rm D}^{\alpha, {\rm
    T}}\right)^q \times \nonumber \\
&& \hspace*{0cm} \left({\rm
    [Pi]}/K_{\rm D}^{\alpha,{\rm P}}\right)^p \binom{p}{r} \left({\rm [ADP]}/K^{\alpha, {\rm D}}_{\rm
    D}\right)^r .\label{eq:Z_p_q_r}
 \end{eqnarray}
Here, $\beta$ is the inverse temperature, $E^{\alpha}_{\rm m}$ is the
 energy of an unphosphorylated monomer in state $\alpha$ (with no
 nucleotide bound), and $K_{\rm D}^{\alpha,{\rm s}}$ is the
 dissociation constant for the binding of species ${\rm s}$ to the
 hexamer in state $\alpha$; T denotes ATP, D ADP, and P a phosphate
 group Pi. Since nucleotide exchange is assumed to be
 fast, it is meaningful to integrate over the number of
 nucleotides. This yields the following partition function for a hexamer in state $\alpha$ with
 phosphorylation level $p$:
\begin{eqnarray}
Z^{\alpha} (p) &=& e^{-\beta N E^{\alpha}_{\rm m}} \left(1 +
{\rm [ATP]}/K_{\rm D}^{\alpha, {\rm T}}\right)^{N-p}  \times \nonumber\\
&&\left[{\rm [Pi]}/K_{\rm D}^{\alpha,{\rm P}} \left(1+{\rm [ADP]}/K^{\alpha, {\rm D}}_{\rm D}\right)\right]^p. \label{eq:Z_p}
\end{eqnarray}

The (excess) chemical potential of a hexamer in conformational state
$\alpha$ at phosphorylation level $p$ is given by
\begin{equation}
\label{eq:mu_p}
\mu^{\alpha}(p) = -k_{\rm B} T \ln[Z^{\alpha}(p)].
\end{equation}
We stress that the chemical potential of a KaiC hexamer depends upon the
chemical potentials (the concentrations) of the nucleotides:
$\mu^{\alpha}(p) = \mu^{\alpha}(p; \mu^{\rm T}, \mu^{\rm D}, \mu^{\rm
P})$.  

We will now consider the symmetric model of Fig.1A of the main
text. In this model, the energy levels of the active and inactive
conformational state are mirror images of each other. This is for
reasons of clarity, and not because it is essential. In fact, the full
model of section III is asymmetric, with the active state being more
stable than the inactive one.

\begin{figure}[t]
\center
\includegraphics[width=8cm]{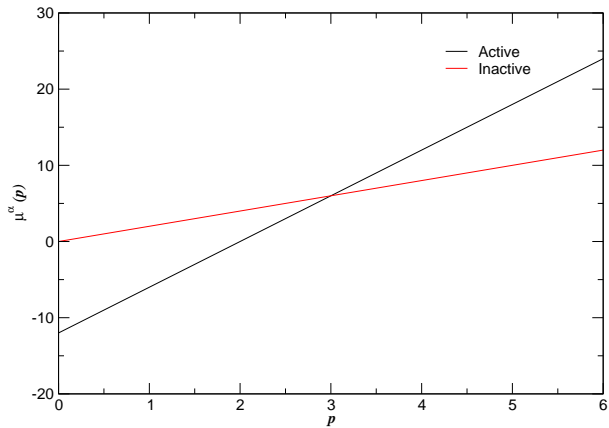}
\caption{\label{fig:FA_I} The chemical potential of a KaiC hexamer as
  a function of the phosphorylation level $p$, for both the active and
  inactive conformational state, and for the symmetric model of Fig.1A
  in the main text. The chemical potential $\mu^\alpha(p)$ is given by
  Eq.~\ref{eq:mu_p}.}
\end{figure}

Fig.~\ref{fig:FA_I} shows the chemical potentials of a KaiC hexamer in
the active and inactive state, respectively, as a function of the
phosphorylation level, for the energy diagram shown in Fig.1A of the
main text. Here, $\mu^{\alpha} (p=6) - \mu^\alpha (p=0)$ corresponds
to the free-energy change upon fully phosphorylating a KaiC hexamer in
conformational state $\alpha$ at constant chemical potentials of the
ATP, ADP and Pi molecules, but {\em without} an ATP hydrolysis
reaction (we thus consider the reaction ${\rm KaiC + Pi} \rightarrow {\rm
  KaiCPi}$) . It is seen that the free energy increases markedly for both
conformational states, meaning that the probability that a hexamer
would fully phosphorylate spontaneously, is essentially zero. Indeed,
the essence of our allosteric model is that in the active
conformational state, the energy from ATP hydrolysis is used to
phosphorylate the KaiC hexamer, while in the inactive state
dephosphorylation occurs spontaneously.

When ATP is hydrolyzed to $p$-fold phosphorylate a single KaiC hexamer
in the active state, the total change in free energy of the system is
\begin{equation}
\Delta G^{\rm A} (p) = \mu^{\rm A} (p) - \mu^{\rm A}(0) + p (\mu^{\rm D} + \mu^{\rm P} - \mu^{\rm T}),
\label{eq:dG_p}
\end{equation}
where $\mu^{\rm A} (p)$ is given by Eq.~\ref{eq:mu_p}. 
If in the active state  binding of both ADP and ATP is strong
 (i.e.  ${\rm [ADP]/K_{\rm D}^{\rm A,ADP}}, {\rm [ATP]/K_{\rm
D}^{\rm A,ATP}} \gg 1$), then the above expression reduces to
\begin{eqnarray}
\Delta G^{\rm A} (p) &=& p (-\Delta G^{\rm A,T} + \Delta G^{\rm A,P} + \Delta G^{\rm A,D} + \Delta G_{\rm hydro})\nonumber\\  
 &=& p \Delta G^{\rm A}_{{\rm m; AATP} \rightarrow {\rm A_pADP}}.
\label{eq:dG_A_ATP}
\end{eqnarray}
Here, $\Delta G^{\alpha,s} = + k_{\rm B} T \ln K_{\rm D}^{\alpha,s}$
is the binding free energy of species ${\rm s}$ and $\Delta G_{\rm
hydro}$ is the standard reaction free energy of an ATP hydrolysis
reaction. The overall free energy change $\Delta G^{\rm A}(p)$
corresponds to that of $p$ phosphotransfer reactions ${\rm A}{\rm ATP}
\rightarrow {\rm A_p}{\rm ADP}$ on the active KaiC hexamer (A denotes
a subunit in the active state). The
free-energy change be understood by noting that in the limit of strong
nucleotide binding considered here, the unphosphorylated monomers are
essentially always occupied by ATP, while the phosphorylated monomers
are essentially always occupied by ADP; see also Fig.1A of the main
text.

\begin{figure}[t]
\center
\includegraphics[width=8cm]{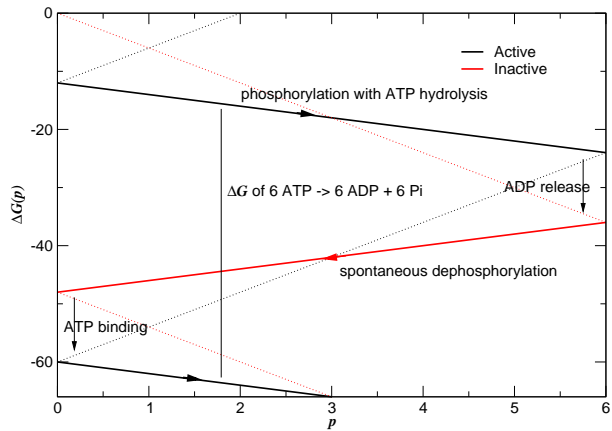}
\caption{\label{fig:FA_ATP_I} The free energy of the system as a
function of the phosphorylation level in the presence of ATP
hydrolysis for the symmetric model of Fig.1A in the main text. The solid lines denote the path of the system. Driven by
ATP hydrolysis, a KaiC hexamer is phosphorylated in the active
state. When the KaiC hexamer is (nearly) fully phosphorylated, it
flips from the active to the inactive conformational state. In the
inactive state, ADP is released and the hexamer dephosphorylates
spontaneously. At low phosphorylation levels, the hexamer flips back
to the active state. The active hexamer rebinds ATP and the
phosphorylation cycle starts over again. The dotted lines correspond
to driven phosphorylation of inactive KaiC and spontaneous
dephosphorylation of active KaiC.}
\end{figure}

Fig.~\ref{fig:FA_ATP_I} shows the free energy of the system in the
presence of ATP hydrolysis. In the active conformational state, KaiC
binds ATP. ATP hydrolysis then drives the phosphorylation of the
hexamer, and the reduction in free energy of the whole system is given
by Eq.~\ref{eq:dG_A_ATP}. When the hexamer is (nearly) fully
phosphorylated, it flips to the inactive conformational state. In the
inactive state, nucleotide binding is weak and, as a result, ADP is
released. The hexamer now dephosphorylates spontaneously; since
nucleotide binding is weak, the free-energy change is given by
$6k_{\rm B} T \ln\left[{\rm [Pi]}/K_{\rm D}^{\rm I,P}\right]$ (see
Eqs.~\ref{eq:Z_p} and~\ref{eq:mu_p}). At low phosphorylation levels,
the inactive hexamer flips back towards the active conformational
state. KaiC rebinds ATP and the phosphorylation cycle starts over
again. After one full phosphorylation cycle, the free energy of the
system has been reduced by the free-energy change corresponding to $6$
ATP hydrolysis reactions: $\Delta G = 6 \left (\Delta G_{\rm hydro} +
k_{\rm B} T \ln\left({\rm [ADP][Pi]}/{\rm [ATP]}\right)\right)$.


\subsection{Dynamics: a transition-state theory of the conformational transitions}
So far, we have discussed the thermodynamics of the phopshorylation
cycle. We will now discuss the {\em dynamics} of the cycle, in
particular the dynamics of the conformational transitions. This is
important, because while the large amplitude oscillations as observed
experimentally require that the hexamers should not flip at
intermediate phopshorylation levels, the stability of one
conformational state with respect to that of the other, does change in
sign at intermediate phosphorylation levels: in the symmetric model
considered here and in Fig.1A of the main text, the active state is
more stable for $p<3$, while the inactive state is more stable for
$p>3$. How can we explain that the conformational transitions
predominantly occur when the hexamers are either nearly fully
phosphorylated or fully unphosphorylated?

This is a difficult question to answer, because it requires knowledge
of the microscopic dynamics of the transition paths between the
conformational states. However, if we assume that nucleotide binding
is an important component of the reaction coordinate that
describes the conformational transitions, then we can make an estimate
of the flipping rates using a mesoscopic model based on concepts
from transition-state theory~\cite{Chandlerbook}.

If nucleotide binding contributes to the reaction coordinate of the
conformational transitions, then we cannot integrate it out as we have
done so far. To derive the flipping rates, we start by considering the
free-energy difference between two conformational states with the same
number of nucleotides bound:
\begin{equation}
\Delta G(p,q,r) = N \Delta E_{\rm m} + p \Delta E_p + q \Delta E_{\rm T} + r \Delta E_{\rm D}.
\end{equation}
Here, $\Delta E_{\rm m} = E_{\rm m}^{\rm A} - E_{\rm m}^{\rm I}$ and
$\Delta E_{\rm s} = k_{\rm B}T \ln K^{\rm A,s}_{\rm D}/K^{\rm
I,s}_{\rm D}$. If we assume, for simplicity, that the model is
symmetric, $\Delta E_{\rm m} = 0$, and that the difference in binding
energy between the active and inactive state is the same for ATP and
ADP, $\Delta E_{\rm T} = \Delta E_{\rm D} = \Delta E_{{\rm T/D}}$,
then the above expression reduces to:
\begin{equation}
\Delta G(p,n) = p \Delta E_p + n \Delta E_{{\rm T/D}},
\label{eq:dG_sym}
\end{equation}
where $n$ is the number of nucleotides that are bound.  We iterate
that phosphorylation favors the inactive state, and hence $\Delta
E_{\rm p} > 0$, while nucleotide binding favors the active state,
$\Delta E_{\rm T/D} < 0$. We can use the above expression to estimate the
flipping rate if we assume that nucleotide binding is the dominant
reaction coordinate for the flipping process.

\begin{figure}[t]
\center
\includegraphics[width=8cm]{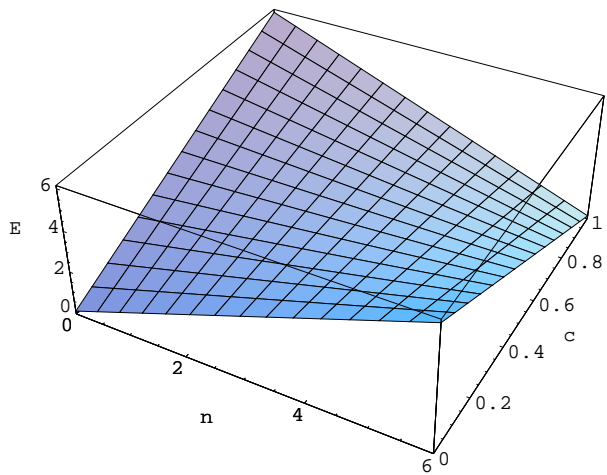}
\caption{\label{fig:E_n_c} The free energy of a KaiC hexamer for a
  phosphorylation level of $p=3$, as a function of the number of
  bound nucleotides $n$ and as a function of an order parameter $c$ that
  denotes the conformational state of the KaiC hexamer: it is zero if
  the hexamer is in the inactive state and one if it is in the active
  state. In the inactive state, essentially no nucleotides are bound,
  and $n \approx 0$, while in the actives state, because of strong
  nucleotide binding, $n\approx 6$. It is seen that in order to flip
  from the inactive to active state, the system has to cross a
  free-energy barrier; the transition state is denoted by the red cross. We imagine that the height of the barrier to go
  from the inactive to active state is given by the free-energy to add
  three nucleotides, while the barrier to flip from active to inactive
  is given by the free energy to remove 3 nucleotides.}
\end{figure}

\begin{figure}
\vspace*{0.3cm}
\center
\includegraphics[width=8cm,angle=0]{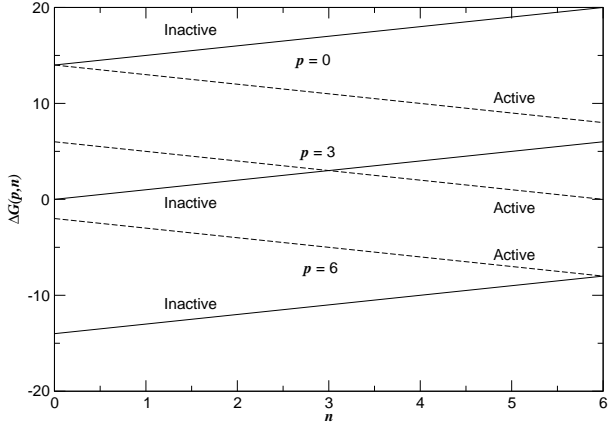}
\caption{\label{fig:dG_p_n} The free energy of the active and inactive
  state as a function of the number of bound nucleotides, $n$, for
  three different phosphorylation levels, $p=0, 3, 6$.  In the active
  state, $n \approx 6$, while in the inactive state, $n \approx
  0$. The free energy of $p$-fold phosphorylated hexamer in state
  $\alpha$ with $n$ nucleotides bound is given by $\Delta G^{\alpha}
  (p,n) = -k_{\rm B}T\ln \sum_{q,r} Z^\alpha (p,q,r) \delta(q+r-n)$,
  where $Z^\alpha(p,q,r)$ is given by Eq.~\ref{eq:Z_p_q_r}.}
\end{figure}

This is illustrated in Figs.~\ref{fig:E_n_c} and~\ref{fig:dG_p_n}, for
the case $|\Delta E_p| = -|\Delta E_n|$. Fig.~\ref{fig:E_n_c} shows a
sketch of the free-energy surface $\Delta G(p=3,n,c)$ of a three-fold
phosphorylated hexamer as a function of the number of bound
nucleotides, $n$, and as a function of an order parameter that
describes the conformational state of the hexamer, $c$; the parameter
$c$ is zero if the hexamer is in the inactive state and one if it is
in the active state. Clearly, we do not know what would be the best
order parameter to describe the conformational transition, let alone
what the free energy would be as a function of this order parameter
for different values of $n$. Nevertheless, we do have some knowledge
of the free-energy surface: we know how the free energy $\Delta
G(p,n,c)$ changes as a function of $n$ for $c=0$ and for $c=1$ -- this
is given by the free energy of nucleotide binding to the inactive and
active state, respectively; this free energy is related to the log of
the partition function in Eq.~\ref{eq:Z_p_q_r}. We
therefore make the minimal assumption that the free-energy surface
$\Delta G(p,n,c)$ is given by a linear interpolation between the two
functions $\Delta G(p,n,0)$ and $\Delta G(p,n,1)$. This leads to the
surface shown in Fig.~\ref{fig:E_n_c}.

In the active state, nucleotides bind the hexamer very strongly and,
consequently, $n\approx6$, while in the inactive state nucleotides
bind the hexamer rather weakly and $n\approx0$. The two (meta)stable
states of the hexamer are thus the active state with six nucleotides
bound and the inactive state with no nucleotides bound. These two
states are separated by a ``transition-state'' surface: in order to go
from one (meta) stable state to the other, the system has to cross a
free-energy barrier. We assume that the transition state is given by
the saddle-point in the free-energy surface $\Delta G(p,n,c)$, as shown
in Fig.~\ref{fig:E_n_c}. This means that both the location and the height of the free-energy barrier for flipping are determined by  that number $n^*$ for
which the two states become equally stable, $\Delta G(p,n^*) = 0$
(see Fig.~\ref{fig:E_n_c} and Eq.~\ref{eq:dG_sym}). Clearly, the location of the transition
state depends upon the phosphorylation level $p$ of the hexamer: in
the symmetric model considered here, the two conformational states are
equally stable if the number of bound nucleotides is $n^* = p$ (see
Fig.~\ref{fig:dG_p_n} and Eq.~\ref{eq:dG_sym}). The height of flipping from the active to inactive state is thus given by 
\begin{equation}
\beta \Delta G^*_{{\rm A}\rightarrow {\rm I}} (p) =
-\ln\left[\sum_{q,r}Z^{\rm A}(p,q,r) \delta(q+r-p)/Z^{\rm A}(p,0,6)\right],
\end{equation}
while the barrier height for the reverse transition is given by
\begin{equation}
\beta \Delta G^*_{\rm I \rightarrow A} (p) =
-\ln\left[\sum_{q,r}Z^{\rm I} (p,q,r) \delta(q+r-p)/Z^{\rm I}(p,0,0)\right].
\end{equation}
Here, $Z^\alpha(p,q,r)$ is given by Eq.~\ref{eq:Z_p_q_r}.
In words, if a $p$-fold phosphorylated hexamer is in the active state,
with 6 nucleotides bound, then in order to flip to the inactive state
with no nucleotides bound, it has to cross a barrier with a height
that corresponds to the energetic cost of removing $6-p$
nucleotides. Conversely, the height of the barrier for an inactive
hexamer, with no nucleotides bound, to flip to the active state, is
given by the free energy to add $p$ nucleotides. Neglecting entropic
factors, the height of the free-energy barrier thus scales linearly
with the phosphorylation level, leading to the exponential flipping
rates of Eqs.1 and 2 of the main text.

\subsection{Numerical calculations on the allosteric model}
The chemical reactions of the model in Fig.1B of the main text are:

\begin{alignat}{1}
& {\rm C}_i \underset{b_i}{\overset{f_i}{\rightleftarrows}}
  \widetilde{\rm C}_i,\,\,\,
{\rm C}_i \overset{k_{{\rm ps}}}{\rightarrow} {\rm C}_{i+1},\,\,\,
\widetilde{{\rm C}}_i \overset{\tilde{k}_{{\rm dps}}}{\rightarrow}
\widetilde{\rm C}_{i-1}\label{eq:S1} 
\end{alignat}

Here, $C_i$ corresponds to an active KaiC hexamer with phosphorylation
level $i$, while $\widetilde{\rm C}_i$ corresponds to an inactive KaiC
hexamer with phosphorylation level $i$. The first, reversible reaction
corresponds to the conformational transitions of the KaiC hexamers
with forward and backward flipping rates $f_i$ and $b_i$,
respectively, the second corresponds to phosphorylation of active KaiC
at rate $k_{\rm ps}$, while the third reaction corresponds to
dephosphorylation of inactive KaiC at rate $\tilde{k}_{\rm dps}$.

To study the phosphorylation behavior of a {\em single} KaiC hexamer,
we cannot use macroscopic rate equations based on the law of mass
action: these equations would correspond to the average of a {\em
population} of KaiC hexamers. To simulate the behavior of a hexamer, we
have performed kinetic Monte Carlo simulations of the zero-dimensional
chemical master equation corresponding to the reactions in
Eq.~\ref{eq:S1}~\cite{gillespie77}. The solid line in
Fig.1C corresponds to the results of those stochastic simulations.

To study the time evolution of the average phosphorylation level of an
ensemble of KaiC hexamers, we have used macroscopic rate equations based on
the law of mass action. The chemical rate equations that correspond to
Eq.~\ref{eq:S1} are:
\begin{eqnarray}
\frac{d[C_i]}{dt} &=&   \label{eq:SI_CA}  k_{\rm ps} [C_{i\!-\!1}]-  
                        k_{\rm ps} [C_{i}]
                        + b_i  [\widetilde{C}_i] - f_i[C_i]\\
\frac{d[\widetilde{C}_i]}{dt} 
                &=&     \tilde{k}_{\rm dps}[\widetilde{C}_{i\!-\!1}]-
                        \tilde{k}_{\rm dps}[\widetilde{C}_{i}]
                        + f_i  [C_i]
                        - b_i [\widetilde{C}_i]\label{eq:SI_CI}
\end{eqnarray}
The dashed line in Fig.1C of the main text corresponds to the
numerical results of propagating these ordinary differential
equations.

The results in Fig.1C of the main text were obtained with the
following values for the parameters: $k_{\rm ps} = 0.01{\rm hr}^{-1}$,
$\tilde{k}_{\rm dps} = 0.05{\rm hr}^{-1}$, $f_i = 0.1^{N-i}\
{\rm hr}^{-1}$ and $b_i = 0.1^i\ {\rm hr}^{-1}$. We note here that the above
chemical rate equations, corresponding to the model in Fig.1B of the
main text, are identical to those of the simple model with differential
affinity as described in section II with ${\rm [A]_T = 0}$
(see Eqs.~\ref{eq:SI_II_CA}-\ref{eq:SI_II_CI}).  Also the values for
the parameters are indentical to those in
the differential-affinity model of section II (see section below).

\section{Simple Models with Differential Affinity}
In this section, we first provide background information on the
simplified model of the Kai system discussed in section II of the main
text. We then briefly discuss a more generic class of differential
affinity models. 

\subsection{A minimal differential affinity model of the Kai system}
In the simple differential affinity model of section II of the main
text, we assume that only a single KaiA dimer can bind to an active
KaiC hexamer. The chemical reactions of this model are given in
Eqs.3-5 of the main text.  They correspond to the following mass-action kinetic equations for the concentrations $\cact{i}$ and $\cinact{i}$ of KaiC in the active and inactive states, $\ac{i}$ of the KaiA-KaiC complex, and $\acon$ of free KaiA:
\begin{eqnarray}
\frac{d[{\rm C}_i]}{dt} & = & k_{\rm pf}[{\rm AC}_{i-1}] - k^{\rm Af}[{\rm A}][{\rm C}_i] + k_i^{\rm Ab}[{\rm AC}_i] \nonumber \\
& & + \delta_{i,0} b_0 [\widetilde{{\rm C}}_0] - \delta_{i,6} f_6 [{\rm C}_6] \label{eq:secII_complete1}\\
\frac{d[{\rm AC}_i]}{dt} & = & -k_{\rm pf}[{\rm AC}_{i}] + k^{\rm Af}[{\rm A}][{\rm C}_i] - k_i^{\rm Ab}[{\rm AC}_i]  \,\,\,\, (i \neq 6) \\
\frac{d\cinact{i}}{dt} & = & \tilde{k}_{\rm dps}(\cinact{i-1} - \cinact{i}) -\delta_{i,0} b_0 \cinact{0} + \delta_{i,6} f_6 [{\rm C}_6] \\
\frac{d \acon}{dt} & = & -\acon \sum_{i=0}^5 k_i^{\rm Af} \cact{i} +
\sum_{i=0}^5 (k_i^{\rm Ab} + k_{\rm pf}) \ac{i} \; , \label{eq:secII_complete3}
\end{eqnarray}
where $\delta_{i,j}$ is the Kronecker delta, $\tilde{k}_{\rm dps}$ is the spontaneous
dephosphorylation rate, $k_{\rm pf}$ is the rate of
phosphorylation catalyzed by KaiA, and $f_i$ and $b_i$ are
the flipping rates as defined in Eqs.1 and 2 of the main text. The
rates of KaiA binding to and unbinding from active KaiC are respectively
$k^{\rm Af}$ and $k_i^{\rm Ab}$, with the latter dependent on the
number of $i$ of phosphorylated monomers in the KaiC hexamer.  Because we generally
choose parameters such that (un)binding of KaiA to KaiC is much faster
than (de)phosphorylation, it is an excellent approximation to assume
that these binding reactions are equilibrated.  In this case, we do
not have to keep track of $\cact{i}$ and $\ac{i}$ separately; instead,
we obtain dynamical equations for the total concentration $\catot{i} =
\cact{i} + \ac{i}$ of KaiC in the active state and for $\cinact{i}$: 
\begin{eqnarray}
\frac{d\catot{i}}{dt} &=& \label{eq:SI_II_CA} \frac{
                        k_{\rm pf} \acon}
                        {K_{i\!-\!1}\!+\!\acon} \catot{i-1}-
                        \frac{k_{pf} \acon}
                        {K_{i}\!+\!\acon} \catot{i} \nonumber\\
                  &&    + \delta_{i,0} b_i
                        \cinact{i} - \delta_{i,6} f_i \catot{i} \\
 \frac{d\cinact{i}}{dt} &=&
                        \tilde{k}_{\rm
                        dps}(\cinact{i-1} - \cinact{i})
			- \delta_{i,0} b_i
                        \cinact{i} + \delta_{i,6} f_i \catot{i}
			\label{eq:SI_II_CI}
\end{eqnarray}
along with a constraint equation giving the free KaiA concentration $\acon$ implicitly:
\begin{eqnarray}
\label{eq:Aeq1}
\acon+\sum_i \frac{\acon \catot{i}}{K_{i}+\acon}  &=& \acon_{\rm T}.
\end{eqnarray}
 Here $\acon_{\rm T}$ is the total concentration of KaiA and $K_i =
 k_i^{\rm Ab}/k^{\rm Af}$ is the dissociation constant for KaiA
 binding to $i$-fold phosphorylated KaiC in the active state.  We implement differential affinity by setting $K_i =
\beta \alpha^{i}$ with $\alpha > 1$. Here, and in the next section, the differential
equations were solved using Matlab.  We note that the assumption that
 KaiA binding is fast is convenient for some purposes but not
 essential; we have verified that the model's predictions are the same
 whether we use
 Eqs.~\ref{eq:secII_complete1}--\ref{eq:secII_complete3} or the
 reduced set Eqs.~\ref{eq:SI_II_CA}--\ref{eq:Aeq1}.

PARAMETER VALUES GO HERE (or probably, better yet, in a table).

\subsection{Parameter dependence and bifurcation behavior}

It is natural to ask how the model's behavior changes as the
parameters are varied from the specific values listed in the previous
subsection**OR: IN THE TABLE**.  This task is facilitated by its
simple cyclic structure, which allows one to prove that the system of
equations~\ref{eq:secII_complete1}--\ref{eq:secII_complete3} always
admits exactly one fixed point.  As the total concentration
$\acon_{\rm T}$ of KaiA is increased from zero, this fixed point
becomes unstable through a supercritical Hopf bifurcation.  The
resulting stable limit cycle persists through a fairly broad range of
$\acon_{\rm T}$ values before finally disappearing at a second
supercritical Hopf bifurcation where the unique fixed point regains
its stability.  This behavior can be understood based on arguments
similar to those put forth in Section IIIC of the main text:  If $\at$
is too large, then the concentration of KaiA is no longer limiting,
and differential affinity cannot act to synchronize oscillations.  On
the other hand, if $\at$ beocmes too small (while the other parameters
are held fixed), then the phosphorylation
reactions will proceed too slowly compared to the dephosphorylation
reactions.  The first phosporylated hexamers will then be dephosphorylated
and return to state ${\rm C}_0$ before the remaining KaiC complexes
reach state ${\rm C}_6$; the hexamers in state ${\rm C}_0$ will win
the competition for the limited KaiA, preventing the synchronized
release of the remainder of the KaiC.  Varying other parameters has
similar effects.  For example, as the
parameter $\beta$ in $K_i = \beta \alpha^i$ is increased, the limit cycle
will eventually collapse into the fixed point in a Hopf bifurcation.
Indeed, if the $K_i$ become too large, then KaiC hexamers cannot
efficiently bind all of the available KaiA, and differential affinity
is no longer possible.  Figure~\ref{fig:simple_bifurc} shows a
2-parameter bifurcation diagram as a function of $\at$ and $\beta$.
We will see that similar bifurcation behavior reappears in more
realistic models of the Kai system.

\subsection{Generic differential affinity model}
The simple model discussed above and in Section II of the main text
can be seen as one example of a more generic class of models that
use the same mechanism to synchronize oscillations.  Here, we briefly
discuss this broader perspective on differential affinity.  A fuller
mathematical analysis will appear in a forthcoming publication.

We begin by considering the following cycle:
\begin{equation}
{\rm C}_0 \overset{k_0}\rightarrow {\rm C}_1 \overset{k_1}\rightarrow
\ldots \rightarrow {\rm C}_{N-1} \rightarrow
{\rm C}_N \overset{k_N} \rightarrow {\rm C}_0,
\label{eq:abstractcycle}
\end{equation}
where ${\rm C}_i$ is a protein that has been $i$-fold covalently
modified.  At least two of the reactions require a catalyst A acting
with Michaelis-Menten kinetics, and those steps that are not catalyzed
by A
are simple first-order reactions.

Suppose that the reactions ${\rm C}_0 \rightarrow {\rm C}_1
\rightarrow \ldots \rightarrow {\rm C}_j$ require the catalyst A (with
$j<N$).  Then, one might imagine that this system will
oscillate if a) the concentration of A is sufficiently low and b) if the
dissociation constants $K_i$ for the binding of A to ${\rm C}_i$
satisfy $K_0 < K_1 < \ldots < K_j$.  These two conditions together
ensure that A first binds to $\plc{0}$ and catalyzes the reaction
$\plc{0} \rightarrow \plc{1}$; only when the concentration of
$\plc{0}$ has dropped almost to zero does A begin to bind to $\plc{1}$
and catalyze the next reaction, and so on until state $\plc{j}$ is
reached.  It turns out, however, that these requirements alone are not
sufficient.  In addition, we must demand that c) the distribution
of arrival times of different C molecules at $\plc{j}$ is not too wide
compared to the average time to travel from $\plc{j}$ back to
$\plc{0}$ and d) that the distribution of arrival times back at $\plc{0}$
is not too broad.  If the former condition, c), does not hold,
then the fastest C molecules will reach $\plc{0}$ while some of the
slower C molecules have still not reached $\plc{j}$.  As we noted in
the previous subsection, oscillations cannot survive such a
situation:  Because A binds most strongly to $\plc{0}$, the fastest C
molecules will siphon away A from the slowest that still need A to
progress around the cycle; these will then slow
down further until all sychronizing effect of differential affinity
has been lost.  Similarly, differential affinity fails when the
arrival of C molecules at state $\plc{0}$ is too spread out.  Then,
there are relatively few C's competing for A molecules at any given
instant, and even those with the highest number $i$ of modifications
can continue upwards towards $\plc{j}$.  Thus, for oscillations to
occur, $N-j$ must be neither too small nor too large.

We have succeeded in finding limit cycles in this model with $j$ as
small as 1 (corresponding to A binding to the two states $\plc{0}$ and
$\plc{1}$) and $N$ as small as 4.  Fig~\ref{} shows oscillations in
such a model with **PARAMETER VALUES HERE**.  Although the
oscillations persist only up to $N = 7$ for these same parameter
values, there appears to be no maximum allowed value of $N$ if the
rate $k_{\rm ps}$ of the transitions $\plc{j} \rightarrow \plc{j+1}
\rightarrow \ldots \rightarrow \plc{N} \rightarrow \plc{0}$ is allowed
to decrease as $1/(N-j)$.  Indeed, in this case, the mean time to
travel from $\plc{j}$ to $\plc{0}$ remains constant, while the
distribution of travel times becomes narrower, which only enhances
oscillations.

One could imagine many variations on the model just described.
For example, we anticipate that for appropriate parameter values,
oscillations will also occur when A binds to and is sequestered by
$\plc{0}$ but is not required for the transition $\plc{0} \rightarrow
\plc{1}$, while still catalyzing the reaction $\plc{1} \rightarrow
\plc{2}$.  One might also consider cases in which differential
affinity acts on two separated blocks of reactions $\plc{0}
\rightarrow \plc{1} \rightarrow \ldots \rightarrow \plc{j}$ and
$\plc{n} \rightarrow \plc{n+1} \rightarrow \ldots \rightarrow
\plc{n+k}$.
%
In each case, the same basic principles of differential
affinity should be at work.

The models of the Kai system in the main text are related to this generic
class of models.  The phosphorylation cycles of both the minimal
differential affinity model of section II and of the full model of the
Kai system in section III have an active branch where KaiC is
phosphorylated, and an inactive branch where KaiC is dephosphorylated.
Also in both models, KaiA catalyzes the phosphorylation reactions on
the active branch. Both ingredients are inspired by experimental
observations; together they give a concrete example of how the
abstract cycle of Eq.~\ref{eq:abstractcycle} might be
implemented. However, the discussion of the generic model above shows
that from the perspective of synchronising the oscillations, there is
no need to make a distrinction between an active and an inactive
branch.  Indeed, the same formalism could be applied to cycles made of
more than two allosteric conformations, or of multiple different sorts
of covalent modifications, or created in any number of other ways.
The differential affinity mechanism thus has the potential to be
generalized far beyond the Kai system.

\section{Full Model of the Kai System}

\subsection{Model: the chemical rate equations}
The chemical reactions that describe the full model of the Kai system
are given by Eqs.7-11 of the main text. Using the law of mass action,
this leads to the following set of macroscopic chemical rate equations
for the concentration of $\cact{i}$, $\ac{i}$, $\cinact{i}$, $[{\rm
    B}_m\widetilde{\rm C}_i]$,
and $[{\rm A}_m{\rm B}_m\widetilde{\rm C}_i]$:

\begin{eqnarray}
\label{eq:supermodel1}
\frac{d[{\rm C}_i]}{dt} &=&   k_{{\rm ps}}\![{\rm C}_{i\!-\!1}]\!+\!
                        k_{{\rm dps}}\![{\rm C}_{i\!+\!1}]
                        \!-\!(k_{{\rm ps}}\!+\!k_{{\rm dps}})[{\rm C}_i]+
                        k_{{\rm pf}} [{\rm AC}_{i-1}]\nonumber\\
                   &&   \!-\!f_i [{\rm C}_i]
                        \!+\!b_i [\widetilde{{\rm C}}_i]
                        -k^{{\rm Af}}_{i}\![{\rm A}][{\rm C}_{i}] + 
                        k^{{\rm Ab}}_{i}\![{\rm AC}_{i}]\\
\frac{d[{\rm AC}_i]}{dt} &=&  k^{{\rm Af}}_{i}\![{\rm A}][{\rm C}_{i}] 
                        - k^{{\rm Ab}}_{i}\![{\rm AC}_{i}] 
                        \!-\!k_{{\rm pf}} [{\rm AC}_{i-1}]\\
\frac{d[\widetilde{{\rm C}}_i]}{dt} 
                &=&     \tilde{k}_{\rm ps} [\widetilde{{\rm C}}_{i\!-\!1}]
                        \!+\!\tilde{k}_{{\rm dps}} [\widetilde{{\rm C}}_{i\!+\!1}]
                        \!-\!(\tilde{k}_{\rm ps}\!+\!\tilde{k}_{{\rm dps}})
                        [\widetilde{{\rm C}}_i]
                        \!+\!f_i [{\rm C}_i]
                        \!-\!b_i [\widetilde{{\rm C}}_i] \nonumber \\
                &&      - \tilde{k}^{{\rm Bf}}_i[{\rm B}]^m
                        [\widetilde{{\rm C}}_i] + 
                        \tilde{k}^{{\rm Bb}}_i 
                        [{\rm B}_m\widetilde{{\rm C}}_i]\\
\frac{d[{\rm B}_m\widetilde{{\rm C}}_i]}{dt} 
                &=&     \tilde{k}_{\rm ps} [{\rm B}_m\widetilde{{\rm C}}_{i\!-\!1}]
                        \!+\!\tilde{k}_{{\rm dps}} 
                        [{\rm B}_m\widetilde{{\rm C}}_{i\!+\!1}]
                        \!-\!(\tilde{k}_{\rm ps}\!+\!\tilde{k}_{{\rm dps}})
                        [{\rm B}_m\widetilde{{\rm C}}_i]
                        \nonumber \\
                 &&     + \tilde{k}^{{\rm Bf}}_i[{\rm B}]^m
                        [\widetilde{{\rm C}}_i]-
                        \tilde{k}^{{\rm Bb}}_i 
                        [{\rm B}_m\widetilde{{\rm C}}_i]
                         \nonumber \\
                 &&       - \tilde{k}^{{\rm Af}}_i[{\rm A}]^m
                        [{\rm B}_m\widetilde{{\rm C}}_i] 
                        + \tilde{k}^{{\rm Ab}}_i 
                        [{\rm A}_m{\rm B}_m\widetilde{{\rm C}}_i]\\
\label{eq:supermodel5}
\frac{d[{\rm A}_m{\rm B}_m\widetilde{{\rm C}}_i]}{dt} 
                &=&     \tilde{k}_{\rm ps} [{\rm A}_m{\rm B}_m\widetilde{{\rm C}}_{i\!-\!1}]
                        \!+\!\tilde{k}_{\rm dps} 
                        [{\rm A}_m{\rm B}_m\widetilde{{\rm C}}_{i\!+\!1}]
                        \!-\!(\tilde{k}_{\rm ps}\!+\!\tilde{k}_{{\rm dps}})
                        [{\rm A}_m{\rm B}_m\widetilde{{\rm C}}_i] 
                        \nonumber \\
                 &&     + \tilde{k}^{{\rm Af}}_i[{\rm A}]^m
                        [{\rm B}_m\widetilde{{\rm C}}_i] 
                        - \tilde{k}^{{\rm Ab}}_i 
                        [{\rm A}_m{\rm B}_m\widetilde{{\rm C}}_i]
\end{eqnarray}

\noindent Here, the concentrations of free KaiA and KaiB, $\rm [A]$ and $\rm
[B]$, are given by:

\begin{eqnarray}
{\rm [A]} &=&   {\rm [A]_T} - \sum_{i=0}^N 
                ([{\rm AC}_i] + m [{\rm A}_m{\rm B}_m\widetilde{{\rm C}}_i])\\
{\rm [B]} &=&   {\rm [B]_T} - \sum_{i=0}^N 
                (m [{\rm B}_m\widetilde{\rm C}_i] + 
                m [{\rm A}_m{\rm B}_m\widetilde{{\rm C}}_i])
\end{eqnarray}

\noindent The phosphorylation and dephosphorylation rates on the active
branch are $k_{\rm ps}$ and $k_{{\rm dps}}$, respectively, and the
flipping rates are $f_i$ and $b_i$. The active state can bind
KaiA with forward and backward rates $k^{\rm Af}_i$ and $k^{\rm
  Ab}_i$, and KaiA can catalyze phosphorylation with the
rate $k_{\rm pf}$. We assume that in the inactive state KaiC can bind
$m\!=\!2$ KaiB molecules with forward and backward rates
$\tilde{k}^{\rm Bf}_i$ and $\tilde{k}^{\rm Bb}_i$, respectively. This KaiB-KaiC
complex can then sequester $m\!=\!2$ KaiA molecules with forward and
backward rates $\tilde{k}^{\rm Af}_i$ and $\tilde{k}^{\rm Ab}_i$.

\subsection{Model: The free-energy difference between the active and inactive
  state of KaiC}
In the presence of only KaiA, KaiC is phosphorylated to a very high
level of $90-95\%$. This requires that the active state of KaiC be
more stable than the inactive one. However, in order for KaiC to
sustain oscillations in the presence of both KaiA and KaiB, KaiC
should be able to flip from the active state to the inactive one at
higher phosphorylation levels. This means that the free-energy
difference $\Delta G (p)$ between KaiC in its active and inactive
conformational states should be strongly negative, but also that it
should rapidly decrease in magnitude as KaiC becomes fully phosphorylated.  A
simple model that captures this is one in which the addition
of each phosphate group decreases the free-energy difference by an
amount $\Delta E_p$, and moreover where the creation of an interface
between a phosphorylated and an unphosphorylated unit also
destabilizes the inactive state by an additional amount $\epsilon$: upon the addition
of the last phosphate group, two such interfaces are removed, leading
to an extra change in the free-energy difference of $2
\epsilon$ in favor of the inactive state. This leads to the following expression for the free-energy
difference between the active and inactive state:
\begin{equation}
\Delta G(p,q,r) = N\Delta E_{\rm m} + p\Delta E_p - \epsilon
\sum_{\langle i,j\rangle} n_i n_j + q\Delta E_{\rm T} + r\Delta E_{\rm
  D}.
\end{equation}
Here, $n_i$ denotes the phosphorylation state of unit $i$ -- $n_i = 1$
if unit $i$ contains a phosphate group and zero otherwise -- and the
sum $\langle i,j \rangle$ includes all nearest neighbor units of
KaiC. For the full model, $\Delta E_{\rm m}, \Delta E_{\rm T}, \Delta
E_{\rm D} < 0$, and $\Delta E_p, \, \epsilon > 0$. For simplicity, we
assume that $\Delta E_p = 2 \epsilon$.  This yields the following
expressions for the transitions between the active and inactive
states: for the forward rate we have $f_i= \delta \gamma^{N-i}$ for
$1\leq i<N$ and $f_i=10\cdot \delta \gamma^{N-i}$ for $i=0,N$ (this is
due to, respectively, the creation or removal of the interfaces
between phosphorylated and unphosphorylated units). The backward rate
$b_i$ is independent of $i$ and much larger than the forward rate
$f_i$, so that the stability of the inactive state is only due to
binding of KaiB. 

\subsection{Setting the parameters}

\begin{table}
\center
\begin{tabular}{|l|r|}
\hline
$k_{\rm ps}$, $\tilde{k}_{\rm ps}$ & 0.025 ${\rm hr}^{-1}$ \\
$k_{\rm dps}$, $\tilde{k}_{\rm dps}$ & 0.4 ${\rm hr}^{-1}$\\
$k_{{\rm pf}}$ & 1.0 ${\rm hr}^{-1}$\\
$f_i$ & $\{10^{-5},10^{-5},10^{-4},10^{-3},10^{-2},10^{-1}, 10\}$ ${\rm hr}^{-1}$\\
$b_i$ & 100 ${\rm hr}^{-1}$\\
$k^{{\rm Af}}_{i}$ & $1.72\!\cdot\!10^6$ ${\rm M}^{-1}{\rm hr}^{-1}$\\
$k^{{\rm Ab}}_{i}$ & $\{10,30,90,270,810,2430,7290\}$ ${\rm hr}^{-1}$\\
$\tilde{k}^{{\rm Bf}}_i$ & $2.97\!\cdot\!10^{12}\times
\{0.01,1, 1, 1, 1, 1, 1\}$ ${\rm M}^{-2} {\rm hr}^{-1}$\\
$\tilde{k}^{{\rm Bb}}_i$ & $1\!\cdot\!10^2\times
\{10, 1, 1, 1, 1, 1, 1\}$ ${\rm hr^{-1}}$\\
$\tilde{k}^{{\rm Af}}$ & $2.97\!\cdot\!10^{18}\times\{0, 1, 100, 100, 1, 0, 0\}$ ${\rm M}^{-2} {\rm hr}^{-1}$\\
$\tilde{k}^{{\rm Ab}}$ & 100 ${\rm hr}^{-1}$\\
${\rm [A]_T}$ & 0.58 $\mu$M\\
${\rm [B]_T}$ & 1.75 $\mu$M \\
${\rm [C]_T}$ & 0.58 $\mu$M \\
\hline
\end{tabular}
\caption{\label{tb:fmtab} List of parameter values for the full model
  in Section III of the main text.}
\end{table}

Using the expressions for $f_i$ and $b_i$ just derived, the full model contains 39 parameters. However, their values are
very much constrained by the large body of experimental data on this
system. We now describe how we have determined the parameters, and how
critical their precise values are for the behavior of the model.  Unless indicated otherwise, the exact values of the parameters
for the full model are summarized in Table~\ref{tb:fmtab}.

\noindent{\bf Concentrations} The concentrations of KaiA and KaiB
dimers are ${\rm [A]_T=0.58} \mu$M and ${\rm [B]_T=1.75} \mu$M,
respectively, and the concentration of KaiC hexamers is ${\rm
  [C]_T=0.58} \mu$M. This corresponds to a concentration ratio of
(KaiA dimers):(kaiB dimers):(KaiC hexamers) = 1:3:1. The corresponding
{\em monomer} concentrations are $1.17 \mu$M KaiA, $3.5 \mu$M KaiB and
$3.5 \mu$M KaiC. Oscillations in phosphorylation have been observed
for these concentrations in the {\em in vitro} experiments of Tomita
{\em et al.} \cite{kageyama06}. However, it
should be noted that for the results in this article the ratios of KaiA
and KaiB to KaiC, $\rm [A]_T/[C]_T$ and $\rm [B]_T/[C]_T$
respectively, are more important than absolute concentrations. Hence,
we often express concentrations in units relative to $\rm [C]_T$.

\noindent{\bf Flipping rates} Following the discussion in the last
paragraph of the previous section on the model for the free-energy
difference between the active and inactive state of the KaiC hexamers,
we next discuss the values of the flipping rates. As discussed in
that paragraph, it is important that: 1) in the absence of KaiB, the
active state has a lower free energy than the inactive one; 2) the
hexamers should not flip at intermediate phosphorylation levels. In
the previous section, we presented a model for the free-energy
difference between the active and inactive state. Here we give the
values of the flipping rates that are consistent with this
free-energy difference, which is requirement 1), and with requirement
2). The forward rate is $f_i= \delta \gamma^{N-i}$ for $1\leq i<N$ and
$f_i=10\cdot \delta \gamma^{N-i}$ for $i=0,N$, with $\gamma=0.1$ and
$\delta = 2 {\rm hr}^{-1}$.  The backward rate $b_i$ is independent of
$i$ and given by $b = 100{\rm hr}^{-1}$. As long as the two
requirements of this paragraph are satisfied, however, the precise values of the flipping rates
are not important for the behavior of the model.

\noindent{\bf KaiC alone and KaiC + KaiB: Spontaneous
  (de)phosphorylation rates} The spontaneous phosphorylation and
dephoshorylation rates are chosen such that the phosphorylation
behavior of KaiC in the absence of KaiA and KaiB and that in the
presence of only KaiB agrees well with experiment (see Fig. 4 of the
main text). This yields the following rates for, respectively, the
spontaneous phosphorylation and spontaneous dephosphorylation
reactions, both for active and inactive KaiC:
$k_{\rm ps}=\tilde{k}_{\rm ps}=0.025 {\rm hr}^{-1}$ and $k_{{\rm dps}}=
\tilde{k}_{{\rm dps}}=0.4 {\rm hr}^{-1}$. The identical
rates for active and inactive KaiC ensure that the phosphorylation
behavior of KaiC in the absence of KaiA and KaiB is the same as that
of KaiC in the presence of KaiB (which stabilizes the inactive
branch). The values of these rate constants are not free and have to
be carefully chosen, because their sum determines the relaxation rate
of the phosphorylation level of KaiC (in the absence of KaiA), while
their ratio determines the steady-state phosphorylation level of KaiC
(in the absence of KaiA).

\noindent{\bf KaiC + KaiA: rates of KaiA-catalyzed phosphorylation
  reactions} The phosphorylation rate of KaiC in the presence of KaiA
is determined by two factors: 1) the binding affinity of KaiA for
KaiC; 2) the rate $k_{\rm pf}$ for the KaiA-catalyzed phosphorylation
reaction. As discussed in the main text, both the mechanism of
differential affinity and temperature compensation require that the
binding affinities of KaiA for KaiC be high. Given these high binding
affinities, the rate at which KaiC is phosphorylated in the presence
of KaiA is mostly determined by $k_{\rm pf}$. This rate is
thus chosen such that the phosphorylation rate of KaiC in the presence of
KaiA agrees with experiment. This gives $k_{\rm pf}=1{\rm
  hr}^{-1}$. This rate constant cannot be freely chosen, because it
directly affects the phosphorylation rate of KaiC in the presence of KaiA.

The mechanism of differential affinity requires that the affinity of
KaiA for KaiC be high, but decrease substantially as the phosphorylation level of
KaiC increases.  Provided that these constraints are satisfied, the
precise values are of less importance. We have chosen the following
expressions for the forward rate $k^{\rm Af}_i$ and backward rate $k^{\rm Ab}_i$: $k^{\rm Af}_i = 1.72 \cdot 10^6 {\rm M}^{-1} {\rm hr}^{-1}$ and $k^{\rm Ab}_i = \beta\alpha^{i} {\rm hr}^{-1}$, with $\alpha=3$ and $\beta=10$.

\noindent{\bf The binding of KaiA and KaiB to inactive KaiC}
The stoichiometries of the complexes of KaiC bound to KaiB and KaiA are not
known.  We assume that inactive KaiC, $\widetilde{\rm C}_i$ binds $m=2$
KaiB dimers, while the complex ${\rm B}_2\widetilde{\rm C}_i$ can sequester
$m=2$ KaiA dimers. The value of $m=2$ is not very critical, as long as
enough KaiA can be sequestered by inactive KaiC the amount of KaiC
available to catalyze phosphorylation is limited.  We find
that for $m>2$ the system also oscillates, although the phase
boundaries, as shown in Fig.6 of the text, do shift. Clearly, more
experiments are needed to resolve the compositions of these complexes.

Temperature compensation requires that the binding affinities of KaiB
for KaiC be high. But as long as this requirement is fulfilled, the
precise values are of less importance. We assume that KaiB binds
inactive KaiC with forward rate $k^{\rm Bf}_i=2.97\!\cdot\!10^{12}\times\{0.01,
1, 1, 1, 1, 1, 1\} {\rm M}^{-2} {\rm hr}^{-1}$ and backward rate 
$k^{\rm Bb}_i=1\!\cdot\!10^2\times\{10, 1, 1, 1, 1, 1, 1\} {\rm hr^{-1}}$. The low affinity of KaiB for $\widetilde{\rm C}_0$ is to ensure that
unphosphorylated hexamers that are in the inactive state can rapidly
flip towards the active state.

As explained in the main text, it is important that the affinity of
KaiA for KaiB is low for KaiB associated with the highly phosphorylated
inactive KaiC hexamers, but increases strongly for KaiB that is bound
to the less-phosphorylated inactive KaiC hexamers; the precise values
of the dissociation constants are not critical. We have chosen the
following forward rate $\tilde{k}^{\rm Af}_i$ and backward rate
$\tilde{k}^{\rm Ab}_i$: $\tilde{k}^{\rm Af}_i =
2.97\cdot10^{18}\times\{0, 1, 100, 100, 1, 0, 0\} $\rm M$^{-2} {\rm
  hr}^{-1}$ and $\tilde{k}^{\rm Ab}_i = 100 {\rm hr}^{-1}$. The low
affinity of KaiA for ${\rm B}_2\widetilde{\rm C}_0$ is to ensure that
unphosphorylated hexamers that are in the inactive state can rapidly
flip towards the active state.

\noindent{\bf Varying the parameters with temperature}
In order to calculate how the oscillation period in our model varies with temperature, we would have to know how the rate constants
would vary with temperature. This would require knowledge of the
activation barrier for each of the reactions, which we do not have. We
can, however, test how sensitive the oscillation period is to changes
in the rate constants. Here, one could choose various strategies: one
could vary each of the rate constants individually, or one could vary
all the rate constants simultaneously, either in a correlated or an
uncorrelated manner. We have performed many such tests, and they all
reveal that the model is very robust to changes in temperature (given
the experimentally observed insensitivity of the phosphorylation rates
to changes in temperature (see main text)). Since the dissociation
constants and the flip rates enter the model in very different ways,
we show in the main text how the oscillations change when we vary
these two groups of parameters.

When we change the dissociation constants of KaiA and KaiB binding to
examine the effects of temperature changes, we simultaneously change
the dissociation constants for KaiA binding, $K_i$ and
$\widetilde{K}_i$, by a factor $C$, and the association and
dissociation rates for KaiB binding, $k^{Bf}_i$ and $k^{Bb}_i$, by
$1/\sqrt{C}$ and $\sqrt{C}$, respectively. Because KaiB (un) binding
is fast, only the {\em ratios} of their rates---the dissociation
constants---matter, and this is a choice that is consistent with
changing the dissociation constants by a factor $C$.

\subsection{Reduced model}
When KaiA binding and unbinding is sufficiently rapid, it is possible
to further simplify Eqs.~\ref{eq:supermodel1}-\ref{eq:supermodel5}. In
this case, we only explicitly take into account the binding of KaiB to
the inactive KaiC hexamers and we assume that both KaiA association to
the active KaiC hexamers and KaiA sequestration by KaiB bound to the
inactive hexamers is very rapid and can be treated as in
chemical equilibrium.  This leads to the following reduced set of
macroscopic chemical rate equations for the total concentration $[{\rm
    C}_i]_{\rm T}$ of KaiC in the active state, both with and without bound
KaiA, the concnetration of KaiC in the inactive state,
$[\widetilde{\rm C}_i]$, and the total concentration
$[{\rm B}_m\widetilde{\rm C}_i]_{\rm T}$ of KaiC in the inactive state with
KaiB bound, both with and without bound KaiA:

\begin{eqnarray}
\label{eq:reducemodel1}
\frac{d[{\rm C}_i]_{\rm T}}{dt} &=&     \sigma^{{\rm
                        ps}}_{i\!-\!1}\![{\rm C}_{i\!-\!1}]_{\rm T}
                        \!+\!\sigma^{{\rm dps}}_{i\!+\!1} [{\rm
                        C}_{i\!+\!1}]_{\rm T}
                        \!-\!(\sigma^{{\rm ps}}_i\!+\!\sigma^{\rm dps}_i)
                        [{\rm C}_i]_{\rm T}
                        \!-\!\sigma^{\rm Ff}_i [{\rm C}_i]_{\rm T}
                        \!+\!\sigma^{\rm Fb}_i [\widetilde{\rm
                        C}_i] \\
\frac{d[\widetilde{\rm C}_i]}{dt} 
                &=&     \tilde{k}_{{\rm ps}} [\widetilde{\rm C}_{i\!-\!1}]
                        \!+\!\tilde{k}_{{\rm dps}} [\widetilde{\rm C}_{i\!+\!1}]
                        \!-\!(\tilde{k}_{{\rm ps}}\!+\!\tilde{k}_{{\rm dps}})
                        [\widetilde{\rm C}_i]
                        \!+\!\sigma^{\rm Ff}_i [{\rm C}_i]_{\rm T}
                        \!-\!\sigma^{\rm Fb}_i [\widetilde{\rm C}_i] \nonumber\\
                &&      - k^{\rm Bf}_i([{\rm B]_T} - m
                        \textstyle\sum_i \displaystyle
                        [{\rm B}_m\widetilde{\rm C}_i]_{\rm T})^m 
                        [\widetilde{\rm C}_i]
                        + \frac{k^{\rm Bb}_i \widetilde{\rm K}_i^m 
                        [{\rm B}_m\widetilde{\rm C}_i]_{\rm T}}
                        {\widetilde{\rm K}_i^m+[{\rm A}]^m}\\
\frac{d[{\rm B}_m\widetilde{\rm C}_i]_{\rm T}}{dt} 
                &=&     \tilde{k}_{{\rm ps}} [{\rm B}_m\widetilde{\rm
                        C}_{i\!-\!1}]_{\rm T}
                        \!+\!\tilde{k}_{{\rm dps}} 
                        [{\rm B}_m\widetilde{\rm C}_{\!+\!1}]_{\rm T}
                        \!-\!(\tilde{k}_{{\rm ps}}\!+\!\tilde{k}_{{\rm dps}})
                        [{\rm B}_m\widetilde{\rm C}_i]_{\rm T}                         \nonumber \\
                 &&     + k^{\rm Bf}_i({\rm [B]_T} - m
                        \textstyle\sum_i \displaystyle
                        [{\rm B}_m\widetilde{\rm C}_i]_{\rm T})^m 
                        [\widetilde{\rm C}_i]
                        - \frac{k^{\rm Bb}_i \widetilde{\rm K}_i^m 
                        [{\rm B}_m\widetilde{\rm C}_i]_{\rm T}}
                        {\widetilde{\rm K}_i^m+[{\rm A}]^m}
\label{eq:reducemodel3}
\end{eqnarray}
where the concentration of free KaiA, [A], is given by:

\begin{equation}
[{\rm A}]+\sum_{i=0}^N \frac{[{\rm A][C}_i]_{\rm T}}{{\rm K}_{i}+[{\rm A}]} 
+ m \sum_{i=0}^{N} \frac{[{\rm A}]^m[{\rm B}_m\widetilde{\rm
      C}_i]_{\rm T}}
{\widetilde{\rm K}_{i}^m+[{\rm A}]^m} - {\rm [A]_T} = 0 \label{eq:Aeq}
\end{equation}

\noindent The effective (de)phosphorylation rates on the active branch
are given by $\sigma^{{\rm ps}}_i=(k_{\rm ps} {\rm K}_i+k_{\rm pf}[{\rm A}])/({\rm K}_i+[{\rm A}])$ and $\sigma^{{\rm dps}}_i={\rm K}_i k_{{\rm dps}}/({\rm K}_i+[{\rm A}])$. The dissociation constants ${\rm K}_i$ and $\widetilde{\rm K}_i$ are given by ${\rm K}_i=k^{{\rm Ab}}_{i}/k^{{\rm Af}}_{i}$ and $\widetilde{\rm K}_i=\tilde{k}^{{\rm Ab}}_{i}/\tilde{k}^{{\rm Af}}_{i}$. The effective flipping
rates are given by $\sigma^{Ff}_i=f_iK_i/(K_i+[A])$ and $\sigma^{Fb}_i=b_i$, where $f_i$ and $b_i$ are the forward and backward flipping rates. We have
confirmed that for sufficiently large $k^{\rm Af}, k^{\rm Ab}$ and $\tilde{k}^{\rm Af}, \tilde{k}^{\rm Ab}$ this set of rate equations gives results that are identical to those in Eqs.~\ref{eq:supermodel1}-\ref{eq:supermodel5}. Unless indicated otherwise, the results in the main text are obtained by numerically solving Eqs.~\ref{eq:reducemodel1}-\ref{eq:reducemodel3}.

\subsection{Bifurcation analysis}
We have performed a bifurcation analysis of the full model. To this
end, we use Eqs.~\ref{eq:supermodel1}-\ref{eq:supermodel5}. However,
the system of differential equations in
Eqs.~\ref{eq:supermodel1}-\ref{eq:supermodel5} obeys the conservation
law $\sum([{\rm C}_i]\!+\![{\rm AC}_i]\!+\![\widetilde{{\rm
    C}}_i]\!+\![{\rm B}_m\widetilde{{\rm C}}_i]\!+\![{\rm A}_m{\rm
  B}_m\widetilde{{\rm C}}_i])\!=\!{\rm [C]_T}$. As a consequence, a
linear stability analysis would always yield at least one eigenvector
with eigenvalue zero, which complicates detection of bifurcation
points. To eliminate the zero eigenvalue associated with the
conservation of KaiC, we express the concentration of one of the KaiC
complexes in terms of the concentrations of the other KaiC
complexes. We have chosen $[{\rm AC}_N]$ to take this role. Thus,
$[{\rm AC}_N]$ is not a separate dynamical variable, but is instead
defined by:

\begin{equation}
{[{\rm AC}_N]} = {\rm [C]_T}\!-\!\sum_{i=0}^{N-1} [{\rm AC}_i]\!-\!\sum_{i=0}^{N}
                ([{\rm C}_i]+[\widetilde{{\rm C}}_i]+
                [{\rm B}_m\widetilde{{\rm C}}_i] + 
                [{\rm A}_m{\rm B}_m\widetilde{{\rm C}}_i])
\end{equation}
Numerical continuation of the fixed points and limit cycles was performed
with the software package XPPAUT~\cite{xppauto}, which incorporates the
numerical continuation routines from AUTO~\cite{auto}.

\begin{figure}[t]
\center
\includegraphics[width=12cm]{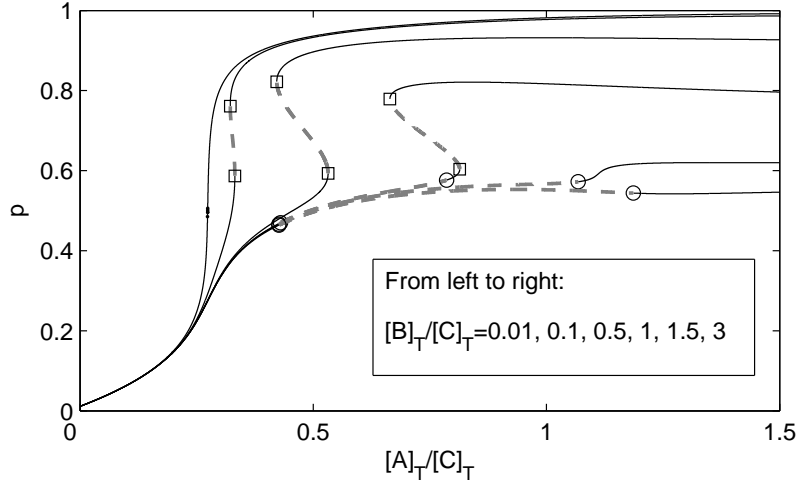}
\caption{\label{fig:BifAna1} Bifurcation diagram of the full model of
  Eqs.~\ref{eq:supermodel1}--\ref{eq:supermodel5} as a function of
  $\rm [A]_T$, for different values of $\rm [B]_T$. Stable fixed
  points and unstable fixed points are indicated by solid black lines
  and dashed grey lines, respectively. When the fixed points change
  stability, either saddle-node bifurcations (squares) or Hopf
  bifurcations (circles) occur. Apart from $\rm [A]_T$ and $\rm [B]_T$, all
  other parameters are as shown in Table~\ref{tb:fmtab}.}
\end{figure}

\begin{figure}[t]
\center
\includegraphics[width=10cm]{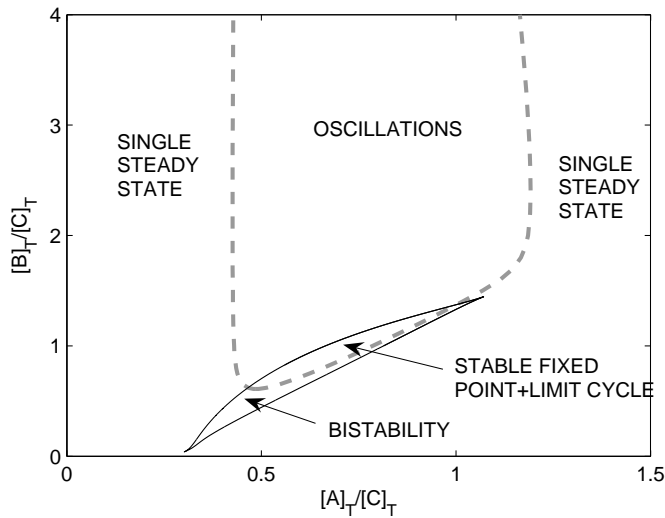}
\caption{\label{fig:BifAna4} Phase diagram of the full model. In the region
enclosed by the dashed grey lines, the system possesses a stable limit
cycle. In the region enclosed by the solid black lines, the system has
three fixed points, of which two are stable in the absence of a limit cycle. Where the two regions overlap, a single stable fixed
point coexists with a limit cycle.}
\end{figure}

In Fig.~\ref{fig:BifAna1} we show the bifurcation diagram of the full
model, defined by Eqs.~\ref{eq:supermodel1}-\ref{eq:supermodel5}, as a
function of $\rm [A]_T$, for different values of $\rm [B]_T$; the
phase diagram is shown in Fig. \ref{fig:BifAna4}. For very small KaiB
concentration, $\rm [B]_T < 0.05 [C]_T$, the system has a single,
stable fixed point for all $\rm [A]_T$ (Fig.~\ref{fig:BifAna1}). For
higher KaiB concentration, $0.05 < \rm [B]_T/[C]_T < 0.6$, the system
is bistable for a range of $\rm [A]_T$ (see Figs. \ref{fig:BifAna1}
and \ref{fig:BifAna4}): it has two stable steady states and one
unstable steady state, corresponding to different degrees of KaiC
phosphorylation. At the boundaries of this bistable region, a stable
and unstable fixed point merge via a saddle-node bifurcation
(Fig. \ref{fig:BifAna1}). We discuss the origin of this bistable
regime in more detail below. For even higher KaiB concentration, $0.6
< \rm [B]_T/[C]_T < 1.2$, one of the two stable fixed points, namely
that with the lower phosphorylation level, becomes unstable for a range of
KaiA concentrations. This stable fixed point becomes unstable via a
supercritical Hopf bifurcation and gives rise to a limit cycle. Thus,
in this window of KaiA and KaiB concentrations, the system has one
stable fixed point at high phosphorylation level and a limit
cycle. For yet larger KaiB concentrations, $\rm [B]_T > 1.2 [C]_T$,
the system has only one unstable fixed point surrounded by a limit
cycle for a range of $\rm [A]_T$; again, the limit cycle appears and
disappears at low and high $\rm [A]_T$, respectively, via a
supercritical Hopf bifurcation. This limit cycle corresponds to the
circadian oscillations discussed in the main
text. Fig. \ref{fig:BifAna4} shows that this oscillatory regime with
only one limit cycle has a lower and an upper bound on the KaiA
concentration, but no apparent upper limit on the KaiB
concentration. In contrast, both the bistable regime and the regime in
which a limit cycle coexists with a stable fixed point, occur
only over a fairly narrow range of KaiA and KaiB concentrations.

\begin{figure}[t]
\center
\includegraphics[width=12cm]{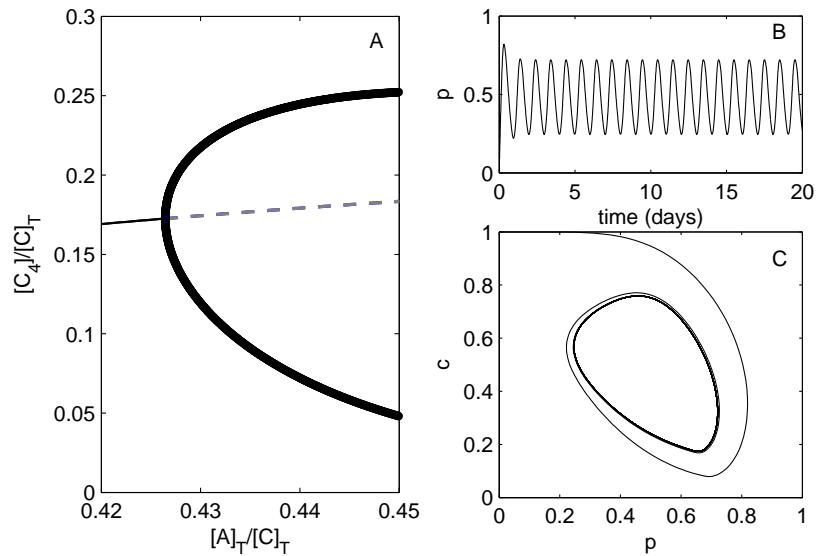}
\caption{\label{fig:BifAna2} Limit cycle in the full model for
$\rm  [B]_T/[C]_T=3$. (A) Bifurcation diagram of $\rm [C_4]$ in the vicinity of the Hopf bifurcation as obtained by numerical continuation of the
limit cycle. The stable and unstable fixed points are indicated by a
solid black line and a dashed grey line, respectively. The minimum and
maximum values of $\rm [C_4]$ along the limit cycle are shown as thick
black lines. The limit cycle is stable, indicating a supercritical
Hopf bifurcation. Here, we choose to plot \rm $[C_4]$ for convenience, and the concentrations of other components of the system show similar behavior
close to the Hopf bifurcation. (B) and (C) Limit cycle for $\rm [A]_T/[C]_T=1$ and $\rm [B]_T/[C]_T=3$, obtained by numerical integration of Eqs.~\ref{eq:supermodel1}-\ref{eq:supermodel5}. (B) Phosphorylation $p$ in time. (C) Phosphorylation $p$ versus $c$, the fraction of KaiC in the active state.}
\end{figure}

In Fig.~\ref{fig:BifAna2}, we examine the properties of the limit
cycle that is created at the Hopf bifurcation. It is possible to do
numerical continuation of the limit cycle in the vicinity of the Hopf
bifurcation as is shown in Fig.~\ref{fig:BifAna2}A.  This analysis
shows that the limit cycle is stable and that the bifurcation is thus supercritical. Further away from the Hopf bifurcation the numerical
continuation algorithm fails to converge. However, as shown in
Fig.~\ref{fig:BifAna2}B and C, by directly solving the differential
equations~\ref{eq:supermodel1}--\ref{eq:supermodel5} we can
nonetheless show that the system continues to converge to a stable
limit cycle.  Because of the fact that the algorithm cannot continue the
limit cycle all the way from one Hopf bifurcation to the other, we
cannot strictly rule out the possibility that it undergoes further
bifurcations. Nevertheless, we find by direct integration of the differential
equations that both the period and amplitude of the limit cycle vary
smoothly with $\rm [A]_T$ between the Hopf bifurcations.

\begin{figure}[t]
\center
\includegraphics[width=12cm]{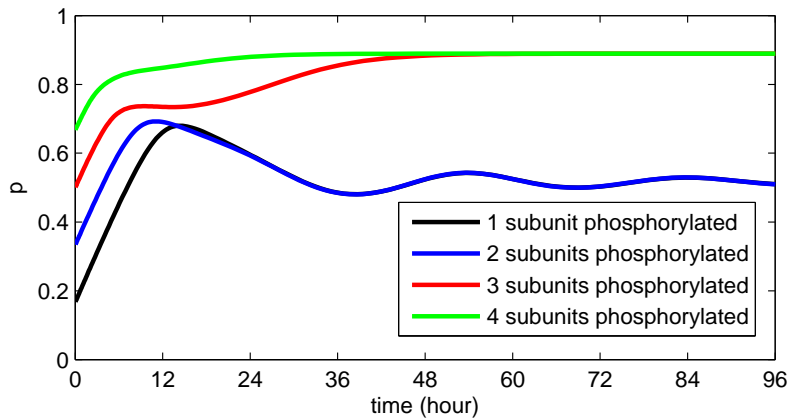}
\caption{\label{fig:BifAna3}
Bistability in the full model for $\rm [A]_T/[C]_T=0.48$ and $\rm [B]_T/[C]_T=0.5$.
The degree of phosphorylation in time is shown for different initial conditions,
$[{\rm C}_i(0)] = {\rm [C]_T}$ for $i=1,2,3,4$. For low initial phosphorylation, the system converges to a steady state at $p\approx0.5$. For high initial phosphorylation the system converges to another steady state at $p \approx 0.9$.}
\end{figure}

We now discuss the origin of bistability for intermediate $\rm [A]_T$
and $\rm [B]_T$ (see Fig. \ref{fig:BifAna4}).  Fig.~\ref{fig:BifAna3}
shows four typical time traces of the phosphorylation level of KaiC
when the system is in the bistable regime. The different time traces
correspond to different initial conditions. These initial conditions
differ in the phosphorylation level of KaiC.
 Indeed, the initial degree of phosphorylation largely
determines which one of the stable fixed points the system converges
to. For low initial phosphorylation, the system converge to a
steady-state phosphorylation level of $p_{\rm s}=0.5$, while for high initial
phosphorylation, it converges to a phosphorylation level of
$p_{\rm s}=0.9$. These two steady states do not only differ in the average
phosphorylation level of KaiC, but, importantly, also in the
concentration of {\em free} KaiA: for $p_{\rm s}=0.5$, $\rm[A]$ is small,
while for $p_{\rm s}=0.9$, $\rm [A]$ is large.

To understand the origin of the difference between the two steady
states, it should be realized that: a) KaiB is needed to stabilize
inactive KaiC, but its concentration is fairly low; b) KaiA is needed
to phosphorylate active KaiC, but also its concentration is fairly
low. 

In the low $p_{\rm s}$ state, most KaiC hexamers have initially a low
degree of phosphorylation. These hexamers will bind KaiA, which will
stimulate their phosphorylation. However, because $\rm [A]_T$ is low,
the phosphorylation rate per hexamer will be low and counterbalanced
by the spontaneous dephosphorylation rate. As a consequence, the
overall phosphorylation level will be low.

In the high $p_{\rm s}$ state, most KaiC hexamers go through the
approximate cycle $\rm AC_5 \rightarrow C_6 \rightarrow
B_2\widetilde{C}_6 \rightarrow B_2\widetilde{C}_5 \rightarrow AC_5$.
In the high $p_{\rm s}$ state, most KaiC hexamers initially have a
high degree of phosphorylation. The available KaiA dimers will be able
to fully phosphorylate these hexamers before they flip towards the
inactive state. On the inactive branch, these hexamers need KaiB to be
stabilized.  However, because $\rm [B]_T$ is low, the inactive
hexamers will not be stabilized very strongly, and will therefore flip
back towards the active state. At this point, the concentration of
free KaiA is still high, because there are no hexamers with a low
phosphorylation level, which could bind KaiA. Because the
concentration of free KaiA is high, the hexamers that have just
flipped back towards the active state can be rephosphorylated, and the
cycle starst again. This situation clearly elucidates the important role of
KaiB. Without KaiB the inactive branch is not stable, and the full
allosteric cycle will be short cut. This will kill the capacity of the
system to generate macroscopic oscillations.

%

The above mechanism for generating bistability is similar to the
mechanisms that have been proposed for generating bistability in the
MAPK \cite{Markevich04} and the CAMKII system \cite{Miller05}. Both in
these systems and in the Kai system, a protein can be phosphorylated
at multiple sites and the limiting amount of either the kinase, as in
the Kai or the MAPK system, or the limiting amount of
phosphatase, as in the CAMKII system, makes the system flip-flop
between a state with a high phosphorylation level and one with a low
phosphorylation level.

Fig.~\ref{fig:BifAna4} summarizes the behavior of the full system. The
full model has a limit cycle for a broad range of
concentrations. Although the range of $\rm [A]_T$ for which
oscillations are observed decreases slightly with increasing $\rm
[B]_T$, we found no indication that oscillations cease for higher $\rm
[B]_T$. This is in agreement with the passive role played by KaiB in
stabilizing the inactive branch and sequestering KaiA. The region for
which bistability occurs is much smaller. Furthermore, as we have
discussed above, the occurrence of the bistable regime does depend
upon details of the model, such as the extent to which KaiB sequesters
KaiA and stabilizes the inactive state when bound to KaiC. Further
experiments will be needed to determine whether bistability really
occurs in the Kai system.

\end{document}